# LASER-PLASMA ACCELERATION IN A CONICAL PLASMA CHANNEL WITH LONGITUDINALLY INHOMOGENEOUS PLASMA PROFILE

*D. S. Bondar[1], W. Leemans[2], V. I. Maslov[1,2], and I. N. Onishchenko[1]*

[1]*National Science Center "Kharkiv Institute of Physics and Technology", Kharkiv, Ukraine*
[2]*Deutsches Elektronen-Synchrotron DESY, Hamburg, Germany*
*E-mail: bondar.ds@yahoo.com*

Laser-plasma acceleration is considered as a modern method of accelerating bunches using a wakefield excited by a laser pulse. This paper demonstrates the use of a longitudinally inhomogeneous increasing plasma density gradient in a conical channel to increase of the energy of a self-injected bunch. Comparison of a conical channels with homogeneous and inhomogeneous plasma and also conical and cylindrical homogeneous channels, shows a clear advantage of an inhomogeneous conical channel. The longitudinally inhomogeneous plasma helps to maintain the self-injected bunch in the wakefield acceleration phase and increases the accelerating gradient. The conical geometry prevents laser pulse expanding, and compress it. The combined effect was shown: the inhomogeneous plasma use, the effect of a conical geometry led to significant increasing the accelerating gradient and longitudinal momentum of the bunch.

## INTRODUCTION

Laser wakefield acceleration (LPA) has emerged as modern method for generating high-energy electron beams in compact setups, overcoming the limitations imposed by conventional accelerators [1, 2, 4].

The fundamental principle of LPA involves an intense laser pulse propagating through plasma, excites the wakefield with strong longitudinal electric field that can accelerate electrons to relativistic energies [1, 3, 4]. As described in [2], these fields can exceed 100 GV/m, orders of magnitude stronger than those achievable in conventional radio-frequency accelerators. This remarkable property has positioned LPA as a promising technology for next-generation compact particle accelerators.

In [5] the review of LPA methods was given; it is highlighted that the self-injection process occurs when plasma electrons are trapped. The self-injected electron bunch parameters are highly sensitive to variations in laser pulse intensity, duration, and plasma density. Only indirect control of self-injected bunch parameters is possible [6, 7]. Over time, the LPA is replaced by a joint LPA and PWFA [8-11].

In [12] authors demonstrated the acceleration of electrons to 1 GeV using a capillary discharge waveguide. The development of quasi-monoenergetic electron beams was another crucial advancement, authors of [13, 14] demonstrating the generation of high-quality beams of relativistic electrons from intense laser-plasma interactions.

Recent investigations have shown that longitudinal inhomogeneous plasma profiles are crucial for controlling self-injection and optimizing acceleration. In [15, 16] authors investigated phase synchronization of self-injected bunches with accelerating wakefield in solid-state plasma, and in [17] this work continued to specifically study wakefield acceleration in inhomogeneous plasma and find self-injected bunches parameters.

The use of plasma density gradient for controlled injection was demonstrated in [18]. Authors showed that plasma-density-gradient injection could produce electron bunch with low absolute-momentum-spread. This technique addresses one of the key challenges in LPA: generating high-quality electron beam. The influence of channel geometry on beam quality has been demonstrated in several studies. Authors of [19] showed that incorporating a hollow plasma channel in a dielectric waveguide significantly enhances radial focusing and improves electron beam quality.

Radially-polarized lasers in parabolic plasma micro-channels can produce MeV-energy electron bunches with low divergence [20]. Corrugated plasma channels enable guiding of laser pulses with subluminal spatial harmonics, allowing for efficient electron acceleration with gradients of several hundred MeV/cm at lower laser powers compared to standard wakefield schemes [21].

High-quality electron bunches can be produced by matching the acceleration length to the dephasing length in laser wakefield accelerators, with cylindrical plasma channel with longitudinally homogeneous and radially parabolic density providing efficient guiding [22].

The corrugated plasma waveguide also enables direct laser acceleration of electrons through quasi-phase matching with radially polarized laser pulses [23]. These techniques demonstrate significant advancements in plasma-based electron acceleration, offering potential for compact, high-energy particle accelerators.

The current study was completed using numerical simulations with the 2D3V code WarpX [24, 25].

## STATEMENT OF THE PROBLEM

As already mentioned in the introduction, the conical plasma channel has advantages over the cylindrical one in holding and focusing the laser pulse.

A similar result was obtained by the authors in [26]. In [27], the longitudinally increasing and decreasing density gradients were experimentally demonstrated.

The increase in density itself has a positive effect on the wakefield amplitude increasing due to the known dependence for the longitudinal acceleration field $E_z \sim \sqrt{n_e}$. The authors of the paper considered the LPA of self-injected bunches in plasma under three cases:
- a homogeneous cylindrical plasma channel (Fig. 1 (a));
- an inhomogeneous conical plasma channel (Fig. 1 (b));
- a homogeneous conical plasma channel (Fig. 1 (c)).

It was used the advantages of a conical channel in focusing of laser pulse [26].

Figure 1 shows the density distribution inside the channel and a graph of the density distribution along the axis (x=0). The authors consider the density increasing in the longitudinal direction according to the law shown on the Fig. 1 (b). The radial profile of the plasma was considered to be homogeneous for all cases.

## RESULTS OF SIMULATION

Fig. 2,3 shows the density graph and the self-injected bunch in the maximum phase of wakefield acceleration.

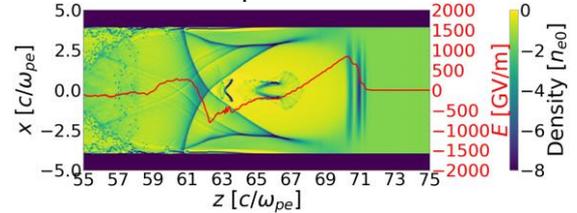

*Fig. 2. Density graph $n_e(z, x)$, acceleration field $E_z(x)$. Cylindrical channel. t=193.4 fs.*

The moment of reaching the end of the channel by the laser pulse is considered (193.4 fs). Detailed profiles of the bunches are shown in Fig. 4, 5. In the case of a cylindrical channel, the self-injected bunch splits (Fig. 4) into two structures.

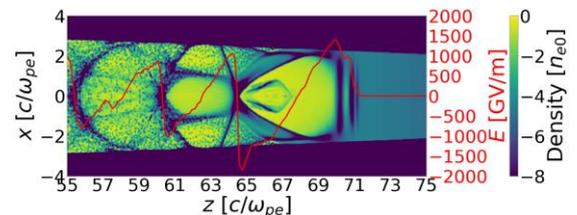

*Fig. 3. Density graph $n_e(z, x)$, acceleration field $E_z(z)$. Conical channel. t=193.4 fs. Inhomogeneous density profile.*

The second self-injected bunch is highly inhomogeneous with a complex structure. Its charge is estimated as small (5.64 pC, I=0.113$I_A$) compared to the first bunch (63.9 pC, I=0.236$I_A$), although at certain points the density is high (up to 5 $n_{e0}$). Charge of self-injected bunch in inhomogeneous conical channel is 75 pC, I=0.3$I_A$. Authors considered first bunch for comparison. In Fig. 5, it can be seen that the self-injected bunch is a whole structure. Similar behavior of self-injected bunches is observed throughout the entire simulation process.

In the case of a cylindrical channel, the formed self-injected bunches are in acceleration fields with an amplitude of 216 GV/m. This is significantly less (5.79 times) than in the case of an inhomogeneous conical channel, in which the acceleration field in the bunches area reaches 1250 GV/m, which is an excellent result for wake acceleration. Fig. 6, 7 and Fig. 8 show the distributions of longitudinal momentum depending on the coordinates (z, x). Interpolation curve is Gaussian. The efficiency of wakefield acceleration is very high in all cases. The value of the total momentum (and energy, taking into account that the self-injected bunches are ultrarelativistic) differs from the longitudinal momentum $p_z$ of about 0.5% and even lower. In this regard, the total momentum value can be considered as energy divided by the light speed.

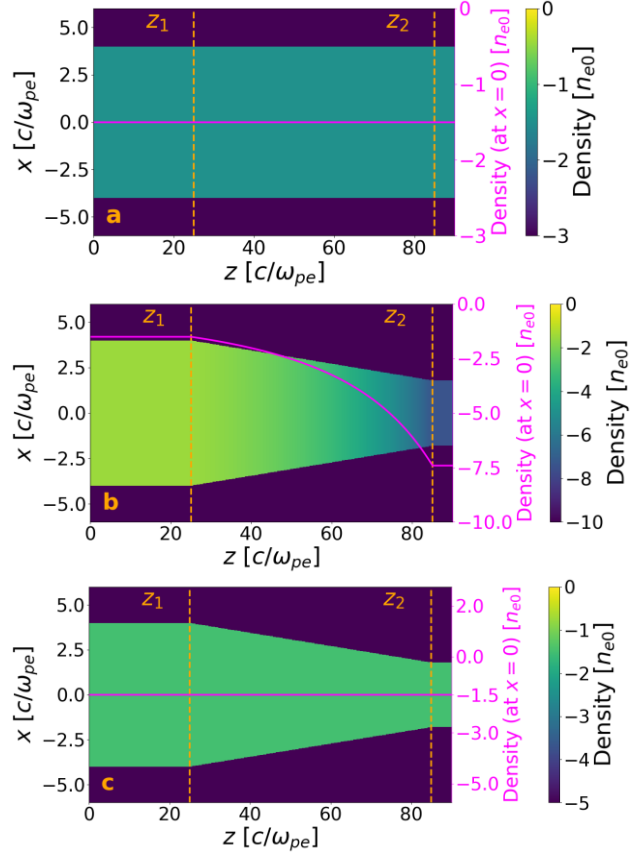

*Fig. 1. Plasma electron density profile $n_e(z, x)$. Longitudinal plasma density profile. Dynamic density change. (a) Cylindrical channel; (b) Conical channel, (c) Conical homogeneous channel. The density outside the channels is 100 $n_{e0}$ (dark violet).*

Thus, the use of a longitudinally increasing density helps to keep self-injected bunch charge and also leads to increase the self-injected bunch longitudinal momentum. Plasma density is normalized to $n_{e0}=1.74 \cdot 10^{19}$ cm$^{-3}$. The wavelength of the laser is $\lambda_{las}$=800 nm. Waist $w_0$=4.95 μm, Full duration $T_{full}$=30.6 fs. Units: $c/\omega_{pe}$=1.27 μm, $1/\omega_{pe}$=4.25 fs. $a_0=eE_0(m_e\omega c)^{-1}$=3. Both the spatial and temporal profiles of the laser are Gaussian. The channel walls are chosen so that they have a density $n_{e,\ walls}$=100$n_{e0}$=1.74$\cdot 10^{21}$ cm$^{-3}$. The bunch current is normalized to the Alfven current $I_A=4\pi\varepsilon_0 m_e c^3 e^{-1}$≈17 kA. This corresponds to the critical density $\omega_{pe}=\omega_{las}$. The length of the simulation area is 90 $c/\omega_{pe}$, the width is 15 $c/\omega_{pe}$. All boundary conditions for particles are absorbing, and for fields are open. A total of $1.57 \cdot 10^6$ macroparticles are considered. In the paper authors compared the cases of a conical channel (inhomogeneous, homogeneous) and a cylindrical channel. In the case of a cylindrical plasma channel, the density was uniform and equal to $n_{e,\ cyl}$=1.5$n_{e0}$.

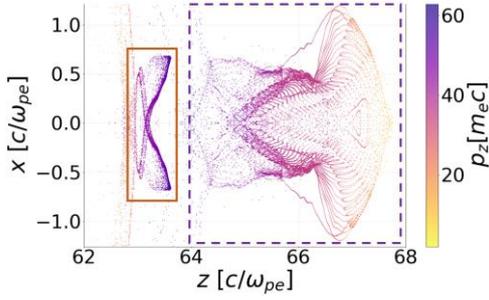

*Fig. 4. Longitudinal momentum $p_z$ of self-injected bunch electrons. Cylindrical channel. t=193.4 fs. The first bunch is highlighted with dots, the second bunch with a solid line.*

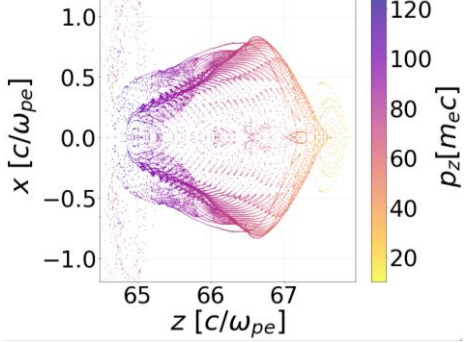

*Fig. 5. Longitudinal momentum $p_z$ of self-injected bunch electrons. Conical channel (inhomogeneous). t=193.4 fs.*

The average value of the longitudinal momentum in the case of a cylindrical channel is $p_z$=36 $m_ec$ for the first bunch, $p_z$=46.47 $m_ec$ for the second. For an inhomogeneous conical channel, it is $p_z$=76.9 $m_ec$ (2.14 times greater than the first bunch and 1.65 times greater than the second bunch). Charge of self-injected bunches: 75 pC, I=0.3$I_A$ (inhomogeneous conical channel), 63.9 pC, I=0.236$I_A$ (first, cylindrical channel), 5.64 pC, I=0.113$I_A$ (second, cylindrical channel). In Fig. 9, it can be seen the electric field distribution in the channel depending on channel profile.

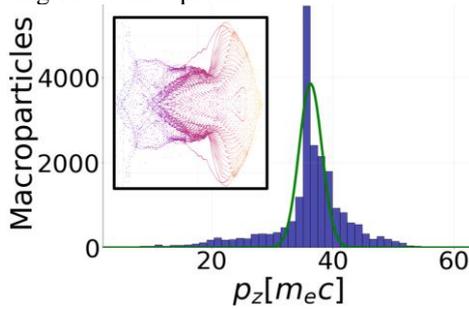

*Fig. 6. Longitudinal momentum $p_z$ distribution. Cylindrical channel. t=193.4 fs (relatively to Fig. 4). First self-injected bunch.*

The conical channel keeps the laser field energy from diverging radially. Similar to the previously obtained result [23], an increase in the energy density on the system axis by 40.9% observed at the conical channel.

Oscillations of the energy density are also observed. Comparison of Fig. 9 (a, b, c) indicates an effective energy transfer from the laser to the acceleration region at the end of the wakefield bubble in the case of a conical channel (the jump in the on-axis EM energy density graph at z=64.5 $c/\omega_{pe}$ in Fig. 9 (b)).

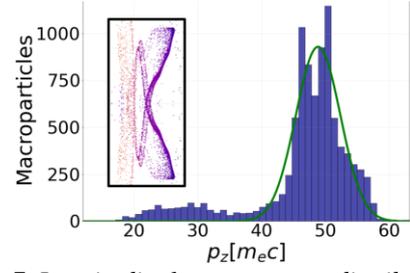

*Fig. 7. Longitudinal momentum $p_z$ distribution. Cylindrical channel. t=193.4 fs (relatively to Fig. 4). Second self-injected bunch.*

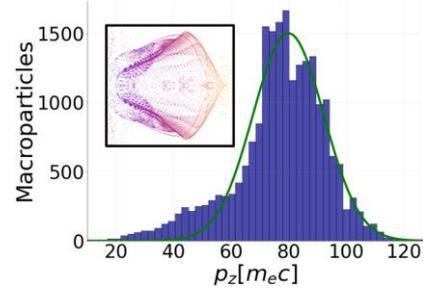

*Fig. 8. Longitudinal momentum $p_z$ distribution. Conical channel (inhomogeneous). t=193.4 fs (relatively to Fig. 5)*

Comparison of the results for an inhomogeneous conical channel (Fig. 9 (b)) and a homogeneous conical channel (Fig. 9 (c)) demonstrates that a conical channel has an advantage in laser focusing and increasing the energy density on the system axis in comparison to homogeneous channel (Fig. 9 (a)). The increase in energy density on the axis in the case of a conical channel was 1.41 times in comparison with cylindrical.

At the same time, compared to a homogeneous conical channel (Fig. 10), the inhomogeneous conical channel (Fig. 3) provides a higher acceleration rate (by 5.79 times) and a higher longitudinal momentum of bunches is by 2.14 times.

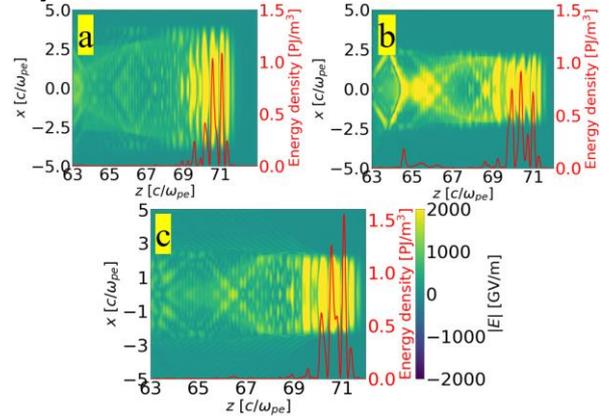

*Fig. 9. Total electric field graph |E| (z, x), on-axis EM energy density (red). (a) Cylindrical channel; (b) Conical inhomogeneous channel, (c) Conical homogeneous channel. t=193.4 fs.*

In the case of inhomogeneous cylindrical channel, in contrast to a homogeneous cylindrical channel (Fig. 2), where charge of the bunch was 63.9 pC, I=0.236$I_A$, in the case of a homogeneous conical channel (Fig. 10), the charge is 44.5 pC, I=0.179$I_A$ (1.44 times lower).

The longitudinal momentum $p_z$ of the self-injected bunch changes from 36 $m_ec$ in the case of a homogeneous cylinder to 44.2 $m_ec$ (by 22.78%) in the case of homogeneous conical channel.

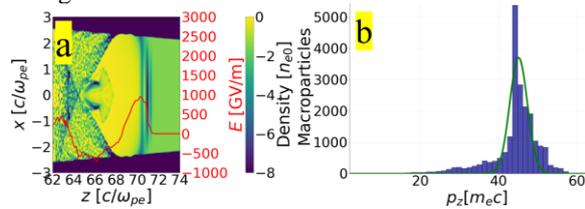

*Fig. 10. (a) Density graph $n_e(z, x)$, acceleration field $E_z(z)$ and (b) $p_z$ distribution. t=193.4 fs.*

Thus, the conical inhomogeneous channel allows obtaining high longitudinal momentum bunches by combining the advantages of inhomogeneous plasma and a channel holding a laser pulse. An increase in the energy density on the axis of the system in the case of a conical channel was also observed.

## CONCLUSIONS

In this paper, a comparison of LPA process in a homogeneous cylindrical channel, and also homogeneous and inhomogeneous conical channels was performed.

A clear advantage of using an inhomogeneous conical channel with plasma density increasing in the longitudinal direction was shown. This leads to an increase in the acceleration field by at least 5.79 times, and the longitudinal momentum of the self-injected bunch by 2.14 times. In addition, focusing of the laser pulse in the conical channel with an increase in energy on the axis was confirmed compared to the cylindrical case. In comparison with the case of a homogeneous conical channel, a significant (almost an order of magnitude) increase in the bunch charge is observed.

Thus, the use of an inhomogeneous conical channel leads to stable acceleration of highly charged and high longitudinal momentum bunches.

## ACKNOWLEDGMENTS

The study is supported by the National Research Foundation of Ukraine under the program "Excellent Science in Ukraine" (project # 2023.03/0182).

## ЛАЗЕРНО-ПЛАЗМОВЕ ПРИСКОРЕННЯ В КОНІЧНОМУ ПЛАЗМОВОМУ КАНАЛІ З ПОЗДОВЖНЬО-НЕОДНОРІДНИМ ПРОФІЛЕМ ПЛАЗМИ

*Д. С. Бондар, В. Ліманс, В. І. Маслов, І. М. Оніщенко*

Лазерно-плазмове прискорення розглядається як сучасний метод прискорення самоінжектованих згустків за допомогою кільватерного поля, збудженого лазерним імпульсом. У цій статті демонструється використання поздовжньо неоднорідного зростаючого градієнта густини плазми в конічному каналі для збільшення енергії самоінжектованого згустку. Порівняння конічних каналів з однорідною та неоднорідною плазмою, а також конічного та циліндричного однорідних каналів вказує на явну перевагу неоднорідного конічного каналу. Поздовжньо неоднорідна плазма допомагає підтримувати самоінжектований згусток у фазі прискорення кільватерної хвилі та збільшує градієнт прискорення. Конічна геометрія запобігає розширенню лазерного імпульсу, стискає його. Було показано комбінований ефект: використання неоднорідної плазми, ефект конічної геометрії призвели до значного збільшення градієнта прискорення та поздовжнього імпульсу згустку.